\documentclass[iop]{emulateapj}

\usepackage{tabularx}
\usepackage{graphicx}
\usepackage[section]{placeins}
\usepackage{array}
\usepackage{amsmath}
\usepackage{amssymb}
\usepackage{wrapfig}
\usepackage{upgreek}
\usepackage{color}
\usepackage{booktabs}
\usepackage{soul}

\def \um {$\upmu$m}

\def \sun {$_{\odot}$}

\def \mnras {MNRAS}
\def \aap {A\&A}

\def \pasp {PASP}
\def \apj {ApJ}
\def \apjs {ApJS}
\def \apjl {ApJL}
\def \pasj {PASJ}
\def \nature {Nature}
\def \jrasc {JRASC}
\def \na {New Ast.}
\def \araa {ARA\&A}

\received{--}
\revised{--}
\accepted{--}

\shorttitle{First results from BISTRO}
\shortauthors{Ward-Thompson et al.}

\begin{document}

\title{First results from BISTRO -- a SCUBA-2 polarimeter survey of the 
Gould Belt}

\author{Derek Ward-Thompson\altaffilmark{1}}
\author{Kate Pattle\altaffilmark{1}}
\author{Pierre Bastien\altaffilmark{2}}
\author{Ray S. Furuya\altaffilmark{3,4}}
\author{Woojin Kwon\altaffilmark{5,6}}
\author{Shih-Ping Lai\altaffilmark{7,8}}
\author{Keping Qiu\altaffilmark{9,10}}
\author{David Berry\altaffilmark{11}}

\author{Minho Choi\altaffilmark{5}}
\author{Simon Coud\'{e}\altaffilmark{2}}
% ...


\altaffiltext{1}{Jeremiah Horrocks Institute, University of Central Lancashire, Preston PR1 2HE, United Kingdom}
\altaffiltext{2}{Centre de recherche en astrophysique du Qu\'ebec \& d\'epartement de physique, Universit\'e de Montr\'eal, C.P. 6128, Succ. Centre-ville, Montr\'eal, QC, H3C 3J7, Canada}
\altaffiltext{3}{Tokushima University, Minami Jousanajima-machi 1-1, Tokushima 770-8502, Japan}
\altaffiltext{4}{Institute of Liberal Arts and Sciences Tokushima University, Minami Jousanajima-machi 1-1, Tokushima 770-8502, Japan}
\altaffiltext{5}{Korea Astronomy and Space Science Institute, 776 Daedeokdae-ro, Yuseong-gu, Daejeon 34055, Republic of Korea}
\altaffiltext{6}{Korea University of Science and Technology, 217 Gajang-ro, Yuseong-gu, Daejeon 34113, Republic of Korea}
\altaffiltext{7}{Institute of Astronomy and Department of Physics, National Tsing Hua University, Hsinchu 30013, Taiwan}
\altaffiltext{8}{Academia Sinica Institute of Astronomy and Astrophysics, P. O. Box 23-141, Taipei 10617, Taiwan}
\altaffiltext{9}{School of Astronomy and Space Science, Nanjing University, 163 Xianlin Avenue, Nanjing 210023, China}
\altaffiltext{10}{Key Laboratory of Modern Astronomy and Astrophysics (Nanjing University), Ministry of Education, Nanjing 210023, China}
\altaffiltext{11}{East Asian Observatory, 660 N. A`oh\={o}k\={u} Place, University Park, Hilo, Hawaii 96720, USA}
% ...

\email{dward-thompson@uclan.ac.uk} 

\label{firstpage}

\begin{abstract} We present the first results from the B-fields In STar-forming Region Observations (BISTRO) survey, using the Sub-millimetre Common-User Bolometer Array~2 (SCUBA-2) camera, with its associated polarimeter (\mbox{POL-2}), on the James Clerk Maxwell Telescope (JCMT) in Hawaii. We discuss the survey's aims and objectives. We describe the rationale behind the survey, and the questions which the survey will aim to answer. The most important of these is the role of magnetic fields in the star formation process on the scale of individual filaments and cores in dense regions. We describe the data acquisition and reduction processes for \mbox{POL-2}, demonstrating both repeatability and consistency with previous data. We present a first-look analysis of the first results from the BISTRO survey in the OMC~1 region. We see that the magnetic field lies approximately perpendicular to the famous `integral filament' in the densest regions of that filament. Furthermore, we see an `hour-glass' magnetic field morphology extending beyond the densest region of the integral filament into the less-dense surrounding material, and discuss possible causes for this. We also discuss the more complex morphology seen along the Orion Bar region. We examine the morphology of the field along the lower-density north-eastern filament. We find consistency with previous theoretical models that predict magnetic fields lying parallel to low-density, non-self-gravitating filaments, and perpendicular to higher-density, self-gravitating filaments.
\end{abstract}



\section{Introduction}

Our knowledge of the star formation process has increased dramatically due to the advent of satellites such as Spitzer and Herschel, and sensitive far-infrared and submillimeter detector arrays such as SCUBA-2. Following on from the highly-successful first-generation JCMT Legacy Surveys, including the Gould Belt Legacy Survey (GBLS; e.g. \citealt{wardthompson2007}; \citealt{buckle2010}; \citealt{graves2010}; \citealt{sadavoy2013}; \citealt{pattle2015}; \citealt{rumble2015}; \citealt{salji2015}; \citealt{chen2016}; Kirk H. et al. 2016\nocite{kirk2016}; \citealt{mairs2016}; \citealt{wardthompson2016}; \citealt{pattle2017}), the JCMT is currently undertaking a series of second-generation surveys, using the latest instruments to be commissioned on the telescope. These include \mbox{POL-2}, an imaging polarimeter for SCUBA-2. One of the surveys using POL-2 is the B-fields in STar-forming Region Observations (BISTRO) Survey that we report here. This is extremely timely because magnetic fields (hereafter referred to as B-fields) are still not well understood in star formation, due to a paucity of observational evidence, despite widespread theoretical recognition of the significance of B-fields in the formation of cores (e.g. \citealt{basu2009} and references therein) and the evolution of proto-stars (e.g. \citealt{li2011} and references therein).

\subsection{Observing magnetic fields}

The submillimeter continuum emission from dust grains is polarised because the grains tend towards alignment perpendicular to B-field lines. For asymmetric particles with some ability to be magnetized, a series of relaxation processes brings the grains towards their lowest energy rotation state. This is with the longest axis perpendicular to the field \citep{lazarian2008}. 

Hence, with material along this axis contributing more to the total far-infrared/submillimeter grain emission, linear polarization is seen perpendicular to the field. In the grain alignment process, the radiative torque that spins up irregularly shaped grains is thought to play the most significant role (e.g. \citealt{lazarian2008}). A few percent polarization is detected astronomically, on scales from proto-stars and jets, up to giant molecular clouds. In some completely symmetric geometries the field lines cancel out so that there is a polarization null. Nevertheless, submillimeter continuum polarization surveys represent a powerful technique for tracing the plane-of-sky B-field orientation (e.g. \citealt{matthews2009}; \citealt{dotson2010}). 

The fractional polarization from dust yields no direct estimate of the B-field strength, since it is dependent on several additional unknowns (e.g., efficiency of grain alignment, grain shape, and composition). However, a measure of the field strength can be derived from the commonly used Chandrasekhar-Fermi (C-F) method \citep{chandrasekhar1953}, and modern variants thereof (e.g. \citealt{hildebrand2009}; \citealt{houde2009}), using dispersion in polarization half-vectors (where high dispersion indicates a highly turbulent velocity field and a weak mean B-field component; `half-vector' refers to the $\pm180$ degree ambiguity in B-field direction), the line widths estimated from spectroscopic data, and the density from the SCUBA-2 flux densities (e.g. \citealt{crutcher2004}; Kirk J. et al. 2006\nocite{kirk2006}). Simulations show that this estimate can be corrected for a statistical ensemble of objects to yield realistic estimates of the field strength (\citealt{ostriker2001}; \citealt{heitsch2001}; \citealt{falceta-goncalves2008}). In addition, the effects of multiple eddies along the line of sight have been studied by \citet{cho2016}.

B-field geometries are generally inferred by preferential emission or absorption by dust or molecules, creating polarized light (e.g., \citealt{cho2007}; \citealt{houde2004}, 2013\nocite{houde2013}). Polarization measurements with molecules require bright lines and are generally restricted to very dense, small-scale structures. Near-infrared absorption polarimetry requires a large sample of background stars and is generally limited to lower-density, more diffuse cloud material (\citealt{goodman1990}; see also \citealt{kwon2015}; \citealt{tamura2015}).

B-field strengths are typically measured using Zeeman splitting of paramagnetic molecules (e.g., \citealt{crutcher2010}).  While detections of Zeeman splitting in the high-density tracer CN have been made towards extremely bright sources (e.g. \citealt{crutcher1996}), Zeeman splitting measurements are typically restricted to lower-density regions of molecular clouds, where the OH molecule is relatively highly abundant (e.g. \citealt{troland2008}).

In contrast, polarized far-infrared and submillimeter thermal dust emission can trace dense structures on both cloud scales and core scales. The Planck satellite has generated an all-sky submillimeter polarization map \citep{planck2015}, allowing us to trace the large-scale B-field over the entire sky. However, it is at too low resolution ($\sim 4$\,arcmin at 857\,GHz; \citealt{planckhfi2011}) to study the detailed cloud geometries in star-forming regions on the necessary scale of prestellar cores and proto-stars.  At somewhat better resolution (30\,arcsec at 250\,\um; \citealt{pascale2008}), the BLASTPol balloon-borne polarimeter has mapped a limited number of star-forming regions in great detail (e.g. \citealt{matthews2014}; \citealt{fissel2015}; \citealt{fissel2016}).

\subsection{Theoretical models}

The theoretical role played by B-fields in star formation has been much discussed (e.g. \citealt{mouschovias1991}; \citealt{mouschovias1991}; \citealt{padoan1999}; \citealt{maclow2004}; \citealt{nakamura2005}; \citealt{vazquezsemadeni2011}; \citealt{inutsuka2015}). However, systematic surveys to measure B-fields in star-forming regions on the necessary resolution scales have proved problematic (see recent reviews by \citealt{crutcher2012}; \citealt{li2014}). \mbox{POL-2} with SCUBA-2 on JCMT is a facility that can map the B-field within cold dense cores and filaments on scales of $\sim$1000-2000~AU in nearby star-forming regions, such as those in the Gould Belt. As such, it can provide a link between the B-field measured on arcminute scales by Planck \citep{planck2015} and BLASTPol (e.g. \citealt{matthews2014}) with measurements made on arcsec scales by interferometers such as the Submillimeter Array (SMA; e.g. \citealt{girart2006}; \citealt{tang2010}; \citealt{chen2012}), Combined Array for Research in Millimeter-wave Astronomy (CARMA; e.g. \citealt{hull2013}; 2014\nocite{hull2014}), and the Atacama Large Millimeter/submillimeter Array (ALMA; e.g. \citealt{nagai2016}; \citealt{cortes2016}). This intermediate size scale is crucial to testing theoretical models of star formation. 

As a result of observations made by the Herschel satellite, it is now widely believed that most low-mass stars form according to the so-called filamentary star formation model \citep{andre2014}. This model has been debated for some time. However, Herschel has shown that this appears to be the dominant star-forming mechanism for solar-type stars \citep{andre2014}. In this scenario a cloud first breaks up into filaments, and material flows onto the filaments along striations, or sub-filaments (e.g. \citealt{palmeirim2013}). A similar picture of movement of material along filaments was previously observed and inferred from a combination of spectroscopic data and simulations (e.g. \citealt{balsara2001} -- using data from \citealt{richer1993}). However, this was just one region. Herschel appears to show the same mechanism in many star-forming regions.

In this model the B-field aligns with the striations (i.e. perpendicular to the filaments), and helps to `funnel' matter onto the filaments.  This observationally-informed paradigm has been reproduced by recent simulations of magnetized self-gravitating filaments (e.g. \citealt{inoue2008}; 2009\nocite{inoue2009}; 2012\nocite{inoue2012}; \citealt{li2010}; \citealt{soler2013}).  Cores then form on filaments, becoming gravitationally unstable and subsequently collapsing to form protostars \citep{andre2014}. 

We know from large-scale polarization studies, e.g. Planck and BLASTPol amonst others (see above) that large-scale fields typically lie roughly perpendicular to their associated filament direction (e.g. \citealt{sugitani2011}; \citealt{palmeirim2013}; \citealt{matthews2014}; \citealt{planck2015}), but we do not know what happens to the field within the dense gas of the filaments themselves, nor what happens within the cores that form in the filaments (c.f BLASTPol; \citealt{matthews2014}). This is crucial to understanding the physical processes taking place, and to discriminating between the models of the star formation process which properly incorporate B-fields (e.g. \citealt{nakamura2005}; \citealt{vazquezsemadeni2011}; \citealt{seifried2015}). 

The current hypotheses are that the field may wrap around the filament in a helical manner (e.g. \citealt{shibata1991} \citealt{fiege2000}); turn to run parallel to the filament in the densest gas (e.g. the purely poloidal field model of \citealt{fiege2000a}); or take on a pinched morphology perpendicular to the long axis of the filament (e.g. \citealt{tomisaka2015}; \citealt{burge2016}), similar to that produced in initially magnetically-supported cores in the classical ambipolar-diffusion paradigm (e.g. \citealt{crutcher2004}; \citealt{galli1993}).

Theoretical studies have shown that both B-fields (e.g., \citealt{li2004}; \citealt{basu2009}) and turbulence (e.g., \citealt{klessen2000}; \citealt{heitsch2011}) can significantly affect how dense structures form, collapse, and evolve in the inter-stellar medium (ISM).  For example, one paradigm of low-mass star formation suggests that collapse is guided by B-fields, producing flattened cores and disks (e.g., \citealt{mouschovias1991}).  This collapse (and subsequent proto-star formation) can drag and twist the field lines, amplifying the local field strength during the early stages of protostellar evolution (e.g., \citealt{machida2005}; \citealt{hennebelle2008a}, \citealt{li2011}).  These twisted lines can then have significant consequences for the emerging protostellar outflows, disks, frequency of binarity, and stellar masses (e.g., \citealt{price2007}; \citealt{hennebelle2008b}; \citealt{machida2011}). 

In fact, there is a debate over the relative importance of B-fields and turbulence in regulating the star formation process (e.g., \citealt{mouschovias1991}; \citealt{padoan2002}). The POL‐2 observations, combined with our existing kinematics from HARP-B (e.g. \citealt{buckle2010}), will allow for an investigation into the balance between gravity, turbulent support, and B-fields, over a statistically meaningful number of star-forming cores in a number of regions across the Gould Belt. 

Once protostars have formed, there is also a debate about the role that the B-field plays in shaping protostellar evolution, and its effect on bipolar outflows. For example, recent studies on the correlation of B-field direction with outflows, using CARMA polarization observations, found no correlation between outflow and field directions on scales below 1000~AU \citep{hull2014}. 

In contrast, a large-scale correlation between outflow and field directions has been found on scales of $\sim$10,000 AU and above \citep{chapman2013}. One explanation of this apparent conflict in the field morphology uses detailed modelling of toroidally-wrapped B-fields at the centres of clouds \citep{seguracox2015}. This has been used to explain early disk formation in Class 0 proto-stars in a recent model in which early disks are hypothesized to preferentially be formed in fields misaligned with the outflow directions \citep{seguracox2015}. \mbox{POL-2} data are crucial to filling in the missing information on intermediate scales between $\sim$1000 and $\sim$10,000 AU. The BISTRO survey aims to address this and all of the other questions discussed above.

Previously, only a few prestellar and protostellar cores have had their B-fields mapped (e.g., \citealt{holland1999}; \citealt{wardthompson2000}; \citealt{matthews2002}; \citealt{wardthompson2009}; \citealt{crutcher2004}; Kirk J. et al. 2006\nocite{kirk2006}). BISTRO will map hundreds. In this paper we describe the plan for the BISTRO survey and discuss the first results taken on OMC~1.

\section{Aims and objectives of the survey}

Previous surveys have either been piecemeal, been very restricted in sample size (e.g., \citealt{matthews2009}; \citealt{vaillancourt2012}; \citealt{hull2013}; 2014\nocite{hull2014}; \citealt{matthews2014}), or have poor resolution to detect cores and proto-stars (e.g. \citealt{planck2015}).  We here describe a project that aims to produce a large and unbiased survey of the B-fields in star-forming molecular material in the solar vicinity, simultaneously at 850 and 450~$\mu$m, and at relatively high resolution -- 14.1 and 9.6~arcsec respectively \citep{dempsey2013}, or $\sim$1000-2000~AU at a typical Gould Belt cloud distance. 

\begin{figure}
\centering
\includegraphics[width=0.47\textwidth]{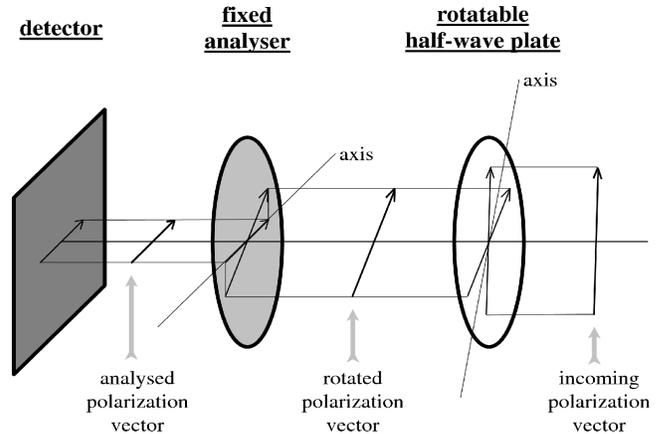}
\caption{A schematic of a rotating half-wave plate polarimeter, from \citet{greaves2003a}.}
\label{fig:fig1}
\end{figure}

The BISTRO Survey is a large-scale survey of the Gould Belt clouds that we have previously mapped in continuum and spectral lines at JCMT (e.g. \citealt{wardthompson2007}; \citealt{buckle2015}; \citealt{white2015}), and in the far-infrared with Herschel \citep{andre2010}.

The aims of the project are: to obtain maps of polarization position angle and fractional polarization in a statistically meaningful sample of cores in numerous regions; to characterize the evidence for and relevance of the B-field and turbulence (in conjunction with previous and follow-up spectroscopic line observations) in cores and their surrounding environments; to test the predictions of low-mass star formation theories (core, filament, outflow, field geometry), and grain alignment theories; to generate a large sample of objects that are suitable for follow‐up with other instruments, such as ALMA, Nobeyama, SMA and NOEMA (NOrthern Extended Millimeter Array); and to measure the B-field strength using the C-F method in as many clouds as possible within our sample.

\begin{figure}
\centering
\includegraphics[width=0.47\textwidth]{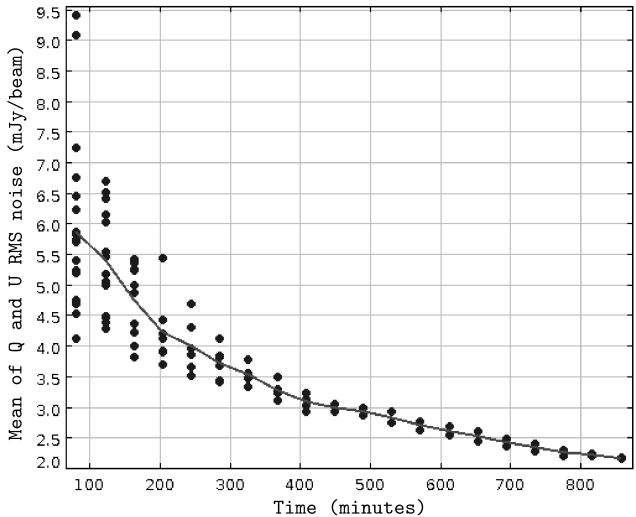}
\caption{A plot of mean measured RMS noise in Q and U, as a function of time (in minutes), at a single off-source position. The data points show the scatter on the individual measurements, and the line is the running mean. A behaviour consistent with t$^{-0.5}$ is seen, as in the ideal case.}
\label{fig:fig2}
\end{figure}

The survey was granted an initial allocation of 224 hours of telescope time to observe 16 fields in 7 different Gould Belt clouds (Auriga, IC5146, Ophiuchus, Orion, Perseus, Serpens and Taurus).  The specific fields were chosen to match those previously mapped by SCUBA-2, HARP and Herschel in the JCMT and Herschel Gould Belt Surveys (\citealt{wardthompson2007}; \citealt{andre2010}).

\section{Observations}
\label{sec:obs}

SCUBA-2 is an innovative 10,000-pixel submillimeter camera \citep{holland2006} that has revolutionized submillimeter astronomy in terms of its ability to carry out wide-field surveys to previously unprecedented depths (e.g., \citealt{buckle2015}; \citealt{pattle2015}). SCUBA-2 uses transition-edge super-conducting (TES) bolometer arrays, which come complete with in-focal-plane super-conducting quantum interference device (SQUID) amplifiers and multiplexed readouts, and are cooled to 100~mK by a liquid-cryogen-free dilution refrigerator \citep{holland2006}.  It has two arrays, which operate simulataneously in parallel, one with filters centered at 850~$\mu$m and one at 450~$\mu$m. In this paper we discuss 850-$\mu$m data only.

\begin{figure*}
\centering
\includegraphics[width=0.75\textwidth]{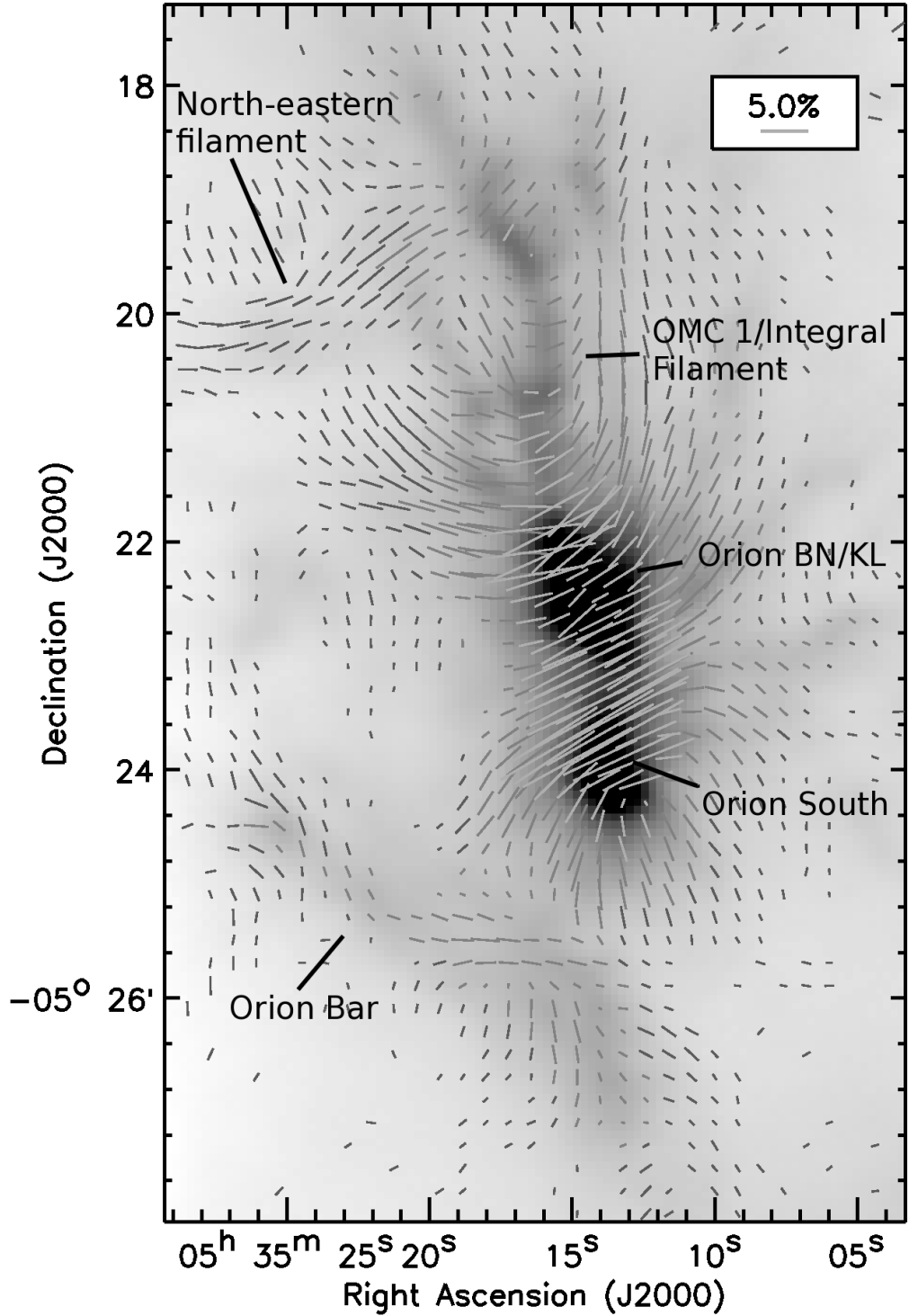}
\caption{A polarization map of the OMC~1 region of the `integral filament' in Orion~A, with half-vectors rotated by 90 degrees to show the B-field direction. The Orion Bar can be seen in the south-eastern part of the map. Half-vectors with $P/DP \geq 3$ are shown. The background image is a SCUBA-2 850-\um\ emission map taken using the standard SCUBA-2 DAISY mapping mode. The half-vector grey-scale is chosen for contrast against the background SCUBA-2 map.}
\label{fig:fig3}
\end{figure*}

The polarimeter \mbox{POL-2} (Bastien et al. 2005a\nocite{bastien2005}; 2005b\nocite{bastien2005a}; 2017; \citealt{friberg2016}) has an achromatic, continuously-rotating, half‐wave plate in order to modulate the signal at a faster rate (2~Hz) than atmospheric transparency fluctuations. Such a modulation improves significantly the reliability and accuracy of submillimeter polarimetric measurements. The signal is analyzed by a wire-grid polarizer. For calibration, a removable polarizer is also available. 

Figure~1 shows a schematic of a rotating half-wave plate polarimeter, such as the POL-2 instrument. POL-2 has three optical components, which are (in the order that the radiation encounters them): the calibration polarizer (not shown in Figure 1), the rotating half-wave plate, and the polarizer. The components are mounted in a box fixed in front of the entrance window of the main cryostat of SCUBA-2. All components are mounted so that they can be taken in and out of the beam remotely, making it very easy and fast to start polarimetry at the telescope (Bastien et al. 2005a\nocite{bastien2005}; 2005b\nocite{bastien2005a}; 2017; \citealt{friberg2016}).

The BISTRO time was allocated to take place during Band~2 weather (0.05 $<$ $\tau_{225~GHz}$ $<$ 0.08), which is typical of moderately good weather conditions on Mauna Kea.  The first data were taken with \mbox{POL-2} on SCUBA-2 on 2016 January 11.

The POL-2 polarimeter fully samples 12-arcmin diameter circular regions at a resolution of 14.1~arcsec in a version of the SCUBA-2 DAISY mapping mode \citep{holland2013} optimised for \mbox{POL-2} observations \citep{friberg2016}.  The \mbox{POL-2} DAISY scan pattern produces a central 3-arcmin diameter region of approximately even, high signal-to-noise ratio coverage, with noise increasing to the edge of the map.  The \mbox{POL-2} DAISY scan pattern has a scan speed of 8~arcsec/sec, with a half-wave plate rotation speed of 2~Hz \citep{friberg2016}.  Continuum observations are simultaneously taken at 450\um\ with a resolution of 9.6~arcsec, but as the 450-\um\ \mbox{POL-2} observing mode has not yet been fully commissioned, we do not use these data in this paper.

The data were reduced in a two-stage process.  The raw bolometer timestreams were first converted to separate Stokes Q and Stokes U timestreams using the process \emph{calcqu} in \textsc{smurf} \citep{chapin2013}.  The Q and U timestreams were then reduced separately using an iterative map-making technique, \emph{makemap} in \textsc{smurf} \citep{chapin2013} and gridded to 4-arcsec pixels.  The iterations were halted when the map pixels, on average, changed by $\leq\,5$ per\,cent of the estimated map RMS noise.  In order to correct for the instrumental polarization (IP), \emph{makemap} is supplied with a total intensity image of the source, taken using SCUBA-2 while \mbox{POL-2} is not in the beam.  The IP correction is discussed in detail by Bastien et al. (2017).  The total intensity image of OMC~1 presented in this paper was taken using the standard SCUBA-2 DAISY observing mode, and reduced using \emph{makemap} using the same convergence criterion and pixel size as the POL-2 data.

The reduced scans were combined in two stages: (1) each of the Stokes Q observations were co-added to form a mosaic Stokes Q image (the Stokes U maps were co-added similarly); (2) each of the Stokes Q and U observations were combined using the process \emph{pol2stack} in \textsc{smurf} \citep{chapin2013} to produce an output half-vector catalogue.  We refer to data produced by this methods as BISTRO Internal Release 1 (IR1).

The data were calibrated in Jy/beam, using an aperture flux conversion factor (FCF) of 725 mJy/pW at 850\um.  When observing with \mbox{POL-2}, the standard SCUBA-2 850-\um\ FCF, of 537 Jy/beam, derived from average values of JCMT calibrators \citep{dempsey2013}, is increased by a factor of 1.35 due to additional losses introduced by \mbox{POL-2} (\citealt{friberg2016}; Bastien et al, 2017).

The OMC~1 region was observed 21 times between 2016 January 11 and 2016 January 25 in a mixture of very dry weather (Band 1; $\tau_{225\,{\rm GHz}}\leq 0.05$) and dry weather (Band 2; $0.05 \leq \tau_{225\,{\rm GHz}}\leq 0.08$) under JCMT project reference numbers M16AL004 (BISTRO) and M15BEC02 (\mbox{POL-2} commissioning).

In order to determine the behaviour of RMS noise in our observations as a function of integration time, we measured the standard deviation on the Stokes Q and Stokes U values in a region with relatively constant signal in both the Stokes Q and the Stoke U maps, located between OMC~1 and the Orion Bar.  This region, centred at approximately $05^{h}35^{m}21^{s}$ $-05^{\circ}23^{\prime}36^{\prime\prime}$ was chosen because it was relatively flat, moderately unpolarised, low in emission, and away from the brightest sources, and because there was no region entirely without signal in the central 3-arcminute-diameter region of the map. Figure 2 shows how the noise integrates down in this 21-repeat ($\sim$14-hour) \mbox{POL-2} observation.

The polarization noise in Figure~2 is seen to integrate down close to t$^{-0.5}$, as in the ideal case. The scatter of individual measurements reduces satisfactorily as the data are subsequently combined. We find that there is no evidence of any `noise floor' in long integrations.  From this plot we see that this dataset has reached 2.1 mJy/beam RMS noise in 13.5 hours. A RMS noise value of $\sim$2~mJy/beam was set as the target value for the BISTRO survey. Appendices A \& B list a series of tests that we carried out to confirm the repeatability of our measurements and to demonstrate consistency with previous data.

\section{First data from the survey}

Figure~3 shows a polarization map taken with POL-2 of the OMC~1 region of the `integral filament' in the Orion~A molecular cloud, with half-vectors rotated by 90 degrees to trace the B-field direction. Only vectors with a signal-to-noise ratio of 3 or greater in polarisation fraction are shown (i.e. $P/DP \geq 3$). The Orion~A molecular cloud is a well-resolved and well-studied region of high-mass star formation (e.g. \citealt{bally2008}; \citealt{odell2008}).  It is the closest region of high-mass star formation, located at a distance of $388\pm 5$\,pc \citep{kounkel2017}.  The half-vector lengths show the percentage polarization, with a 5\% scale bar in the corner to give the calibration.  The underlying image is an 850\um\ total intensity map of the same region taken using SCUBA-2.

The `integral filament' \citep{bally1987} can be seen running roughly north-south through the region. The brightest part of the filament lies just south of the centre of the image. The two brightest and most massive regions in the filament are the northern Becklin-Neugebauer Kleinmann-Low (BN/KL) object (\citealt{becklin1967}; \citealt{kleinmann1967}) and the southern Orion South clump (\citealt{batrla1983}; \citealt{haschick1989}). Both are seen in Figure~\ref{fig:fig3}. In the southeast part of the map the `Orion Bar' photon-dominated region (PDR) extends from the centre of the foot of the map in a roughly northeasterly direction.

In the brightest central part of the filament, the B-field direction, as indicated by the half-vectors, appears to lie roughly orthogonal to the main axis of the filament. This pattern continues on the main axis line of the filament over most of the length of the filament. More particularly, on the brightest part of the filament the orientation of the long axis of the filament is estimated to be $+$11.0 $\pm$ 1.5 degrees, whilst the calculated B-field direction is $-$64.2 $\pm$ 6.5 degrees (both measured north through east; note that there is a 180-degree ambiguity on the B-field direction), yielding a difference of 75.2 $\pm$ 6.7 degrees.  The filament direction was estimated by performing a linear regression on the coordinates of 12 bright peaks of submillimeter emission located along the linear portion of the integral filament, as observed in the JCMT GBS 850\um\ SCUBA-2 data.  The field direction was estimated by taking the mean of the position angles of the B-field half-vectors in the region of uniform field direction in the centre of the OMC~1 region, between the Orion BN/KL and S clumps.

However, away from the central axis of the filament the field appears to curve to either side. In the northern half of the filament the field appears to curve northwards, delineating a roughly `U'~shape, centred on the filament.  In the southern half of the filament the field appears to curve to the south, forming an inverted `U'~shape.  This so-called `hour-glass' morphology was first noted by \citet{schleuning1998} at much lower resolution and signal-to-noise ratio, observing at 100\um\ and 350\um\ with the Kuiper Airborne Observatory (KAO) and the Caltech Submillimeter Observatory (CSO) respectively. However, we note a far higher degree of curvature of the field lines than was seen by \citet{schleuning1998}.

There is a slight degree of de-polarisation visible towards the centres of the BN-KL and Orion-S clumps. This is a well-known effect resulting from tangled fields in the centers of very dense regions (e.g. \citealt{matthews2002}). The pattern along the Orion Bar appears somewhat more complex. Furthermore, in the north-eastern section of the map there is a region of half-vectors that appear to follow a different pattern. Here the half-vectors seem to be running along a different filament. All of the above is consistent with the much lower signal-to-noise-ratio data of \citet{houde2004} and \citet{matthews2009}. The interferometry data of \citet{rao1998} on the peaks of OMC~1 are also consistent with our data. We now discuss all of these features.

\section{Discussion}

Herschel has shown that the dominant formation mechanism for prestellar cores is core formation along filaments \citep{andre2014}, revealing several examples of large-scale filaments lying perpendicular to the (plane-of-sky) B-field directions, as measured with large-scale absorption polarimetry (e.g., \citealt{palmeirim2013}). This is consistent with findings from previous emission polarization measurements from SCUPOL on SCUBA (e.g. \citealt{matthews2001}) and more recent large-scale polarization emission data from BLASTPol (e.g \citealt{matthews2014}).  Based on these examples, a model has emerged whereby collapse occurs first along field lines to form filaments, and then along filaments to form cores \citep{andre2014}. In the lower density regions around the main filament, typically striations (or sub-filaments) are seen parallel to the B-field \citep{palmeirim2013}.

The polarization pattern we have observed in OMC~1 in Figure~3 follows this theoretical picture on-axis. The main part of the integral filament containing the BN-KL object and Orion South has a B-field direction apparently roughly orthogonal to the main filament direction, as mentioned above.

However, our wide-field data also allow us to trace the B-field direction off-axis, and it is here that even more interesting behaviour is seen, as noted above, with a roughly `hour-glass' morphology.  If we follow this theoretical picture, then we would predict that the field lines started out roughly orthogonal to the filament in the lower density as well as the higher density material, in a more uniform configuration, and was subsequently distorted into its current configuration. 

There appear to be two possibilities as to how the hourglass morphology could have formed. One possibility is that the motion of the denser central material along the filament axis pulled the B-field lines into this configuration as predicted by the model (see Figure~9(a) of \citealt{andre2014}). Another possibility is that the well-known BN-KL outflow \citep{thaddeus1972} caused the field lines in the lower density peripheral material to deviate from their original orientation. The effect of the highly-collimated central part of the BN/KL outflow on the B-field on arcsecond scales is discussed by \citet{tang2010}.

We note that the outflow has a wide opening angle, and high-velocity wings with multiple ejecta, often referred to as the `bullets of Orion' \citep{allen1993}.  The central point of the outflow coincides with the position of the BN/KL object, the northern submillimetre-bright region in Figure~\ref{fig:fig3}.  Consequently, the position and opening angle of the outflow roughly match the central part of the hour-glass pattern, as well as the angle between the U-shape and the inverted-U-shape fields, as if the outflow had pushed aside the field.  Further work is required to decide which of these scenarios is correct.

\begin{figure}
\centering
\includegraphics[width=0.47\textwidth]{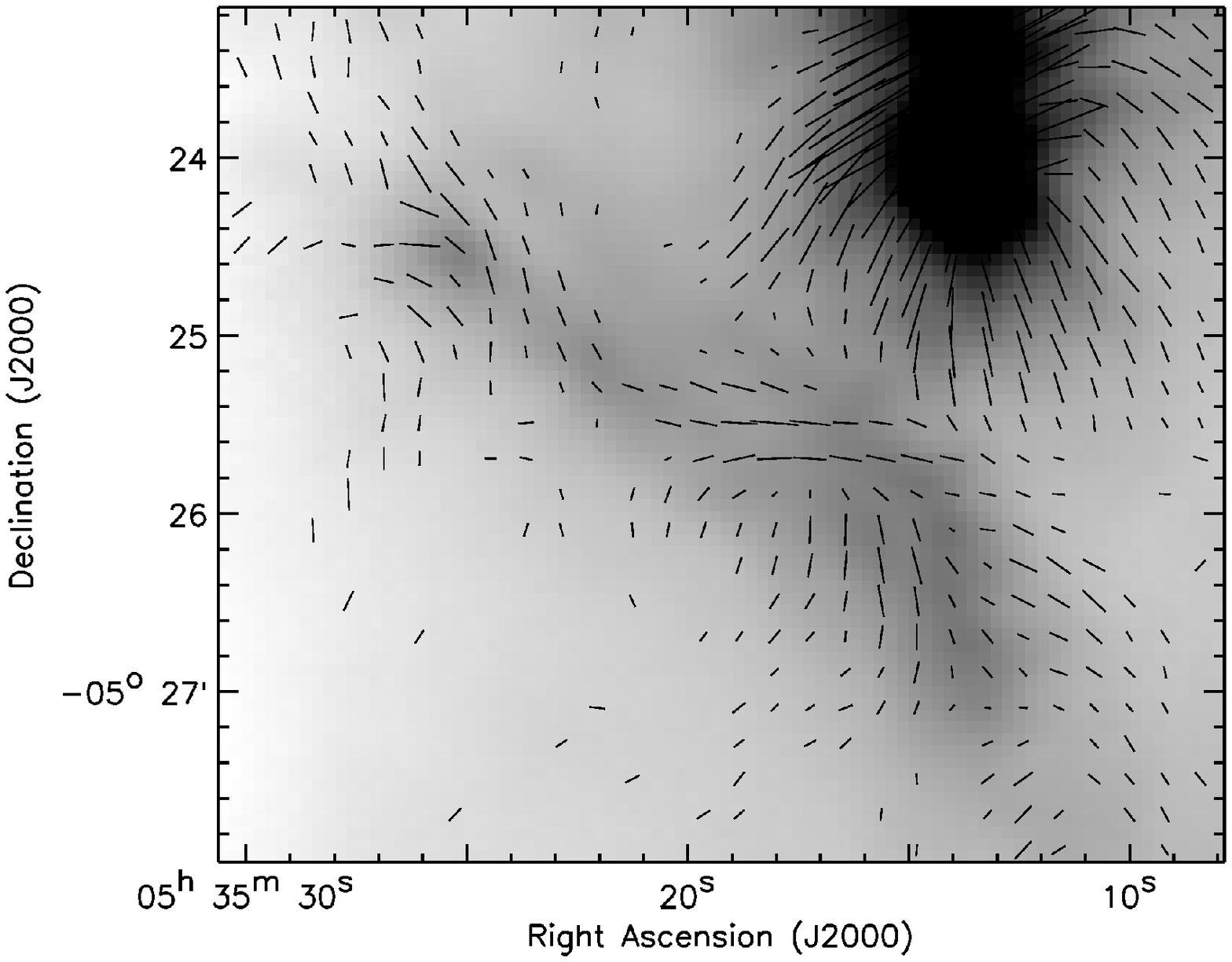}
\caption{The polarization structure of the Orion Bar as observed with POL-2, with half-vectors rotated 90 degrees to show the B-field direction.}
\label{fig:fig4}
\includegraphics[width=0.47\textwidth]{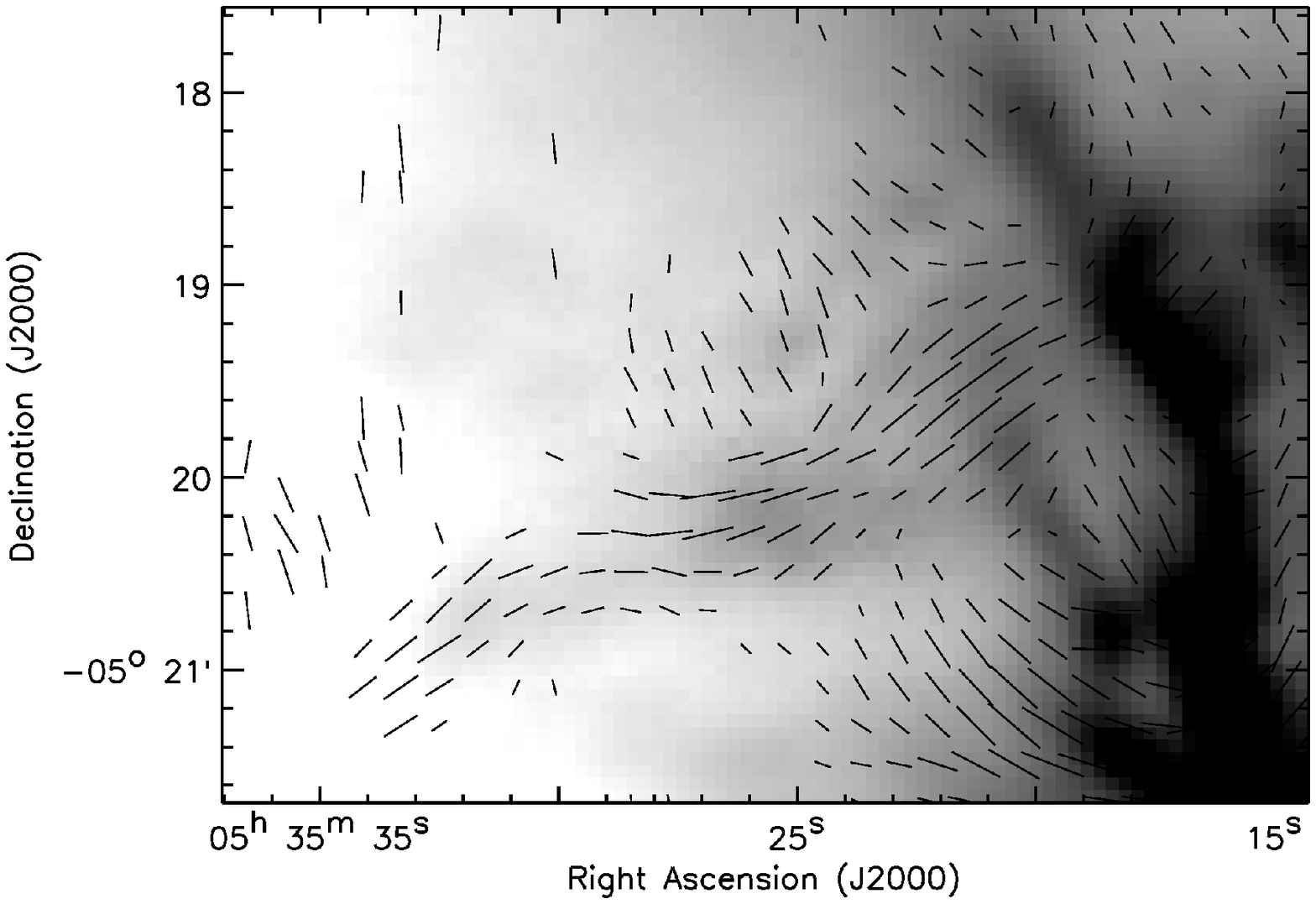}
\caption{The polarization structure of the north-eastern filament as observed with POL-2, with half-vectors rotated 90 degrees to show the B-field direction.}
\label{fig:fig5}
\end{figure}

A close-up of the Orion Bar region is shown in Figure~4. Here we see that the field follows a more complex morphology. At the southern end of the Bar the field appears to be running north-south. In the middle of the Bar the field runs roughly east-west. In the northern part of the Bar the field appears to turn again to run in a north-easterly direction.

This complex pattern clearly indicates a complex field structure.  One possibility is of a field that is simply twisting along the PDR front.  Close examination of the Bar does appear to show the Bar twisting roughly in line with the field direction.  Another possibility is that the field is running helically around the Orion Bar. In such complex cases as this it is often difficult to determine which of a number of different three-dimensional scenarios is being projected onto our two-dimensional field of view (see, e.g. \citealt{franzmann2017}). However, the simulations produced by \citet{franzmann2017} show that a helical field could produce the polarization pattern that we are seeing.

Figure~5 shows a close-up of the north-eastern filament that runs in a roughly east-west direction, and is roughly orthogonal to the main integral filament. This is reminiscent of the sub-filaments, or striations, seen in Taurus \citep{palmeirim2013}, which lie perpendicular to the main filament. Figure~5 shows that the B-field lies roughly parallel to this sub-filament, again as seen in Taurus \citep{palmeirim2013}. Similar behaviour is also seen in the low-density striations in the Polaris Flare region (\citealt{wardthompson2010}; \citealt{panopoulou2016}).

Furthermore, the B-field pattern lying along the north-eastern filament appears to lie in the foreground relative to the hour-glass field. Both north and south of the north-eastern filament the field lies in a direction running northeast-southwest, as if it continues behind the north-eastern filament. Hence, we hypothesise that the north-eastern filament is foreground to the rest of the cloud.

This behaviour of parallel versus perpendicular field geometries is predicted theoretically. For example, numerous studies of non-self-gravitating (i.e. low density) filaments see B-fields lying parallel to filaments -- essentially by running simulations without gravity (e.g. \citealt{ostriker2001}; \citealt{heitsch2001}; \citealt{falceta-goncalves2008}).  \citet{nakamura2008} include self-gravity and see `elongated condensations [i.e. dense filaments] that are generally perpendicular to the large-scale field'.

More recently, \citet{soler2013} studied in detail the effects of varying the B-field strength in a filament, as well as varying the density of the filament. They found that field lines are preferentially perpendicular to the filaments above a certain critical density and parallel to the filaments below this density.  

This is exactly what we see here -- the field is running parallel to the low-density north-eastern filament, and perpendicular to the high-density integral filament (c.f. Figure~1 of \citealt{soler2013}). Incidentally, \citet{soler2013} find field lines perpendicular to filaments only in intermediate-strength and high-strength field cases. This would tend to indicate that the field we are observing in Orion is relatively strong.

\section{Summary}

In this paper we have introduced the BISTRO (B-Fields in STar-Forming Region Observations) survey, which will map the dense regions of many nearby star-forming clouds with the \mbox{POL-2} polarimeter and SCUBA-2 on the JCMT. We have described the rationale behind the survey, and the scientific questions which the survey will answer. The most important of these is the role of B-fields in the star formation process on small scales and in dense regions, and its importance relative to other processes, such as turbulent or non-thermal motions of the gas.

We have described the data acquisition and reduction processes for \mbox{POL-2}, demonstrating that the RMS noise on BISTRO \mbox{POL-2} observations decreases as $t^{-0.5}$ as expected. We presented the first \mbox{POL-2} polarization map from the BISTRO survey, which is of the OMC~1 region of Orion~A, and showed compatibility with previous observations, as well as repeatability of the POL-2 results.

We saw that the field lies perpendicular to the integral filament in the densest regions of that filament. Furthermore, we saw an hour-glass B-field morphology extending beyond the densest region of the integral filament into the less-dense surrounding material, and discussed possible causes for this. We observed a more complex morphology along the Orion Bar. 

We examined the morphology of the field along the lower-density north-eastern filament. We found consistency with previous theoretical models that predict B-fields lying parallel to low-density, non-self-gravitating filaments, and perpendicular to higher-density, self-gravitating filaments.

\section{Acknowledgements}

The James Clerk Maxwell Telescope is operated by the East Asian Observatory on behalf of The National Astronomical Observatory of Japan, Academia Sinica Institute of Astronomy and Astrophysics, the Korea Astronomy and Space Science Institute, the National Astronomical Observatories of China and the Chinese Academy of Sciences (Grant No. XDB09000000), with additional funding support from the Science and Technology Facilities Council of the United Kingdom and participating universities in the United Kingdom and Canada.  Additional funds for the construction of SCUBA-2 and POL-2 were provided by the Canada Foundation for Innovation.  The data taken in this paper were observed under project codes M15BEC02 and M16AL004.  DWT and KP acknowledge Science and Technology Facilities Council (STFC) support under grant numbers ST/K002023/1 and ST/M000877/1.  WK, MK, CWL and SSL were supported by Basic Science Research Program through the National Research Foundation of Korea (NRF), funded by the Ministry of  Science, ICT \& Future Planning (WK: NRF-2016R1C1B2013642; MK: NRF-2015R1C1A1A01052160, SSL: NRF-2016R1C1B2006697) and the Ministry of Education, Science and Technology (CWL: NRF-2016R1A2B4012593).  AP acknowledges the financial support provided by a Canadian Institute for Theoretical Astrophysics (CITA) National Fellowship.  JCM acknowledges support from the European Research Council under the European Community's Horizon 2020 framework program (2014-2020) via the ERC Consolidator grant `From Cloud to Star Formation (CSF)' (project number 648505).  This research has made use of the NASA Astrophysics Data System. The authors wish to recognize and acknowledge the very significant cultural role and reverence that the summit of Mauna Kea has always had within the indigenous Hawaiian community. We are most fortunate to have the opportunity to conduct observations from this mountain.

\textit{Facilities:} James Clerk Maxwell Telescope (JCMT)

\textit{Software:} Starlink \citep{currie2014}, \textsc{smurf} (\citealt{berry2005}; \citealt{chapin2013}), Interactive Data Language (IDL)

\bibliographystyle{aasjournal}

\clearpage

\setcounter{figure}{0}
\setcounter{table}{0}
\renewcommand{\thefigure}{A\arabic{figure}}
\renewcommand{\thetable}{A\arabic{table}}

\begin{figure}
\centering
\includegraphics[width=\textwidth]{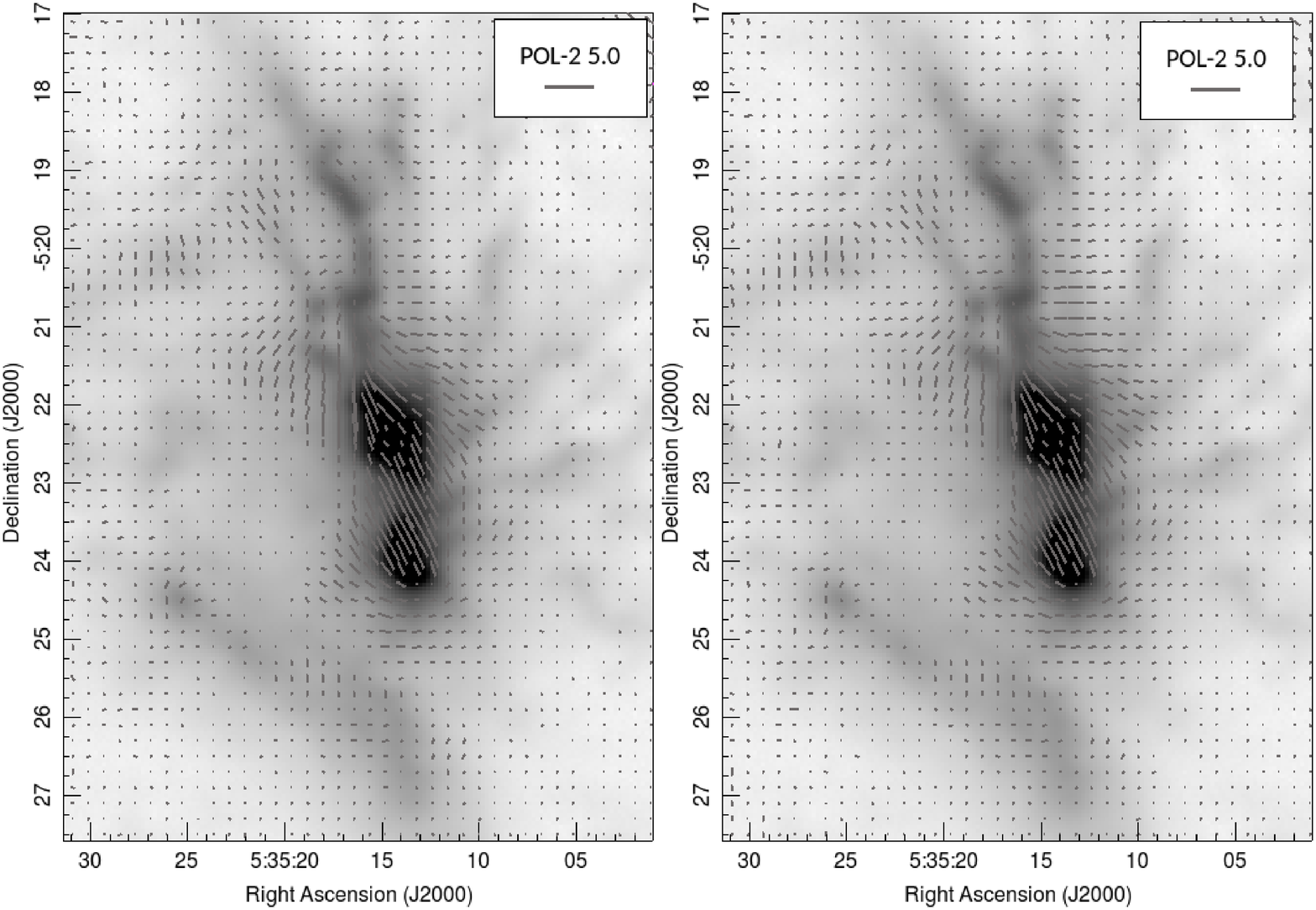}
\caption{Jack-knife test: Polarization maps of the OMC~1 
region of Orion~A made from odd-numbered scans (left) and even-numbered scans (right).  Note that here the half-vectors have not been rotated.}
\label{fig:figa1}
\end{figure}

\section*{Appendix A: Repeatability of \mbox{POL-2} observations}

In this appendix we present a demonstration of the repeatability of 
\mbox{POL-2} observations of extended structure.  These results are a 
subset of a larger study to be presented in the \mbox{POL-2} commissioning 
paper (Bastien et al. 2017), to which we refer the reader for further 
information.

In order to test the repeatability of our observations, we performed 
jack-knife tests on our observations of OMC~1.  We divided the data 
into odd- and even-numbered scans, the half-vector maps produced from which 
are shown in Figure~\ref{fig:figa1}. This division of scans is intentionally arbitrary, and is used to show the 
variation that might be expected between any two samples, uncorrelated in 
any observational property. We see excellent consistency between 
the two maps.

\section*{Appendix B: Comparability of \mbox{POL-2} to previous 
observations}

\setcounter{figure}{0}
\setcounter{table}{0}
\renewcommand{\thefigure}{B\arabic{figure}}
\renewcommand{\thetable}{B\arabic{table}}

\begin{figure}
\centering
\includegraphics[width=0.47\textwidth]{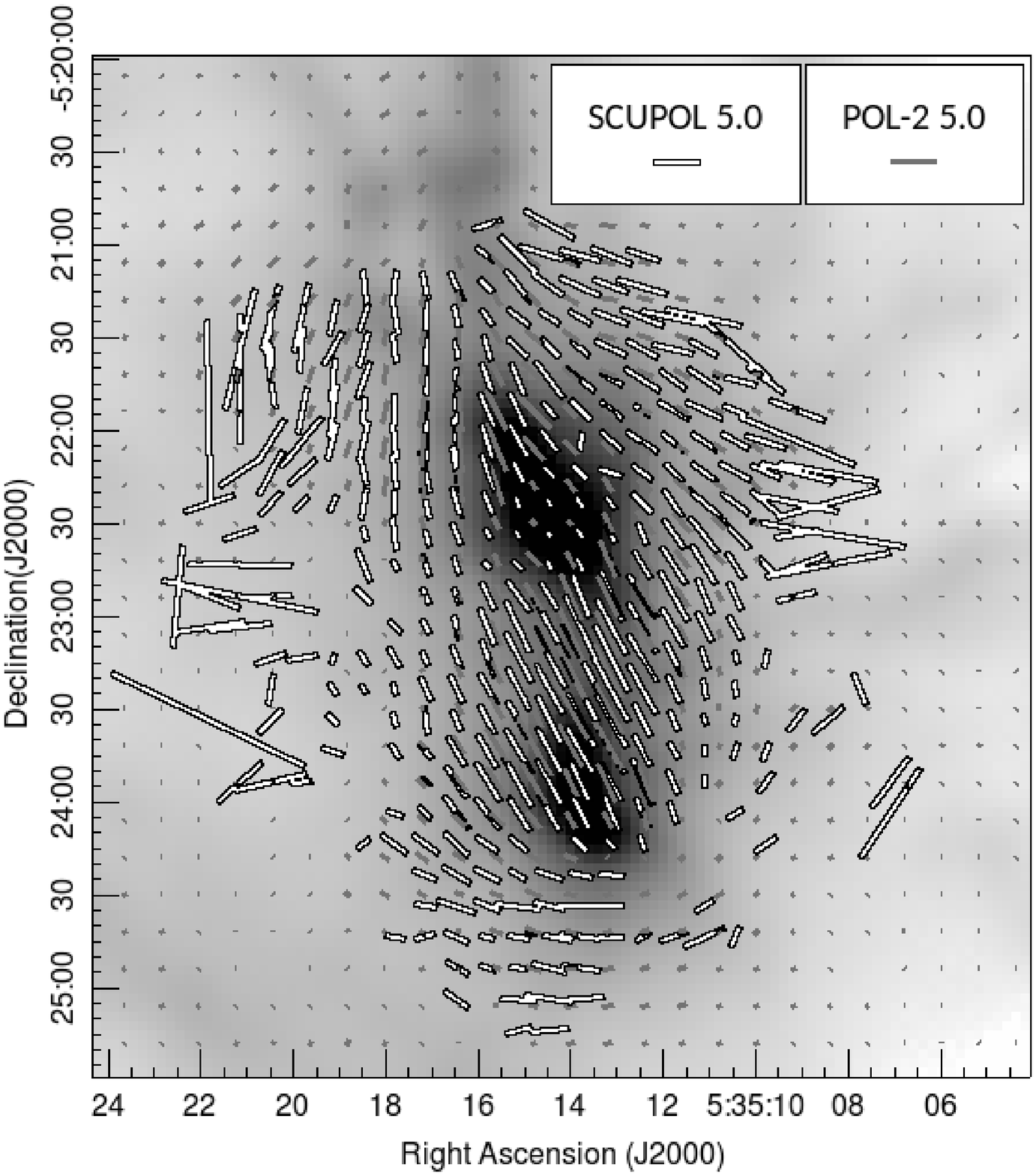}
\caption{The \mbox{POL-2} (grey) and SCUPOL (white) half-vectors, 
overlaid on the JCMT GBS SCUBA-2 image of OMC~1.}
\label{fig:pol2_scupol}
\end{figure}

In this appendix we compare the \mbox{POL-2} map of OMC~1 to previous 
observations of OMC~1 made using the previous JCMT polarimeter, SCUPOL.  
There is no \emph{a priori} reason to expect identical performance from 
SCUPOL and \mbox{POL-2}; the two instruments were/are mounted on different 
cameras (SCUBA and SCUBA-2 respectively; c.f. \citealt{holland1999}; 
\citealt{holland2006}), and take data in different modes 
(c.f. \citealt{greaves2003a}; \citealt{friberg2016}; 
Bastien et al. 2017).  However, the two instruments take data at the 
same wavelength and resolution, and so the data taken ought to be directly 
comparable.

The SCUPOL observations of OMC~1 were published as part of the SCUPOL 
Legacy Catalogue \citep{matthews2009}.  
Figure~\ref{fig:pol2_scupol} shows the SCUPOL data superposed on the POL-2
data. It can be seen that the \mbox{POL-2} and SCUPOL half-vectors 
show a very similar morphology, but that the polarization 
fractions seen in the SCUPOL half-vectors are slightly larger than the 
\mbox{POL-2} half-vectors. We believe that this is due to the lower signal-to-noise
ratio of the older SCUPOL data.

\begin{figure}
\centering
\includegraphics[width=0.47\textwidth]{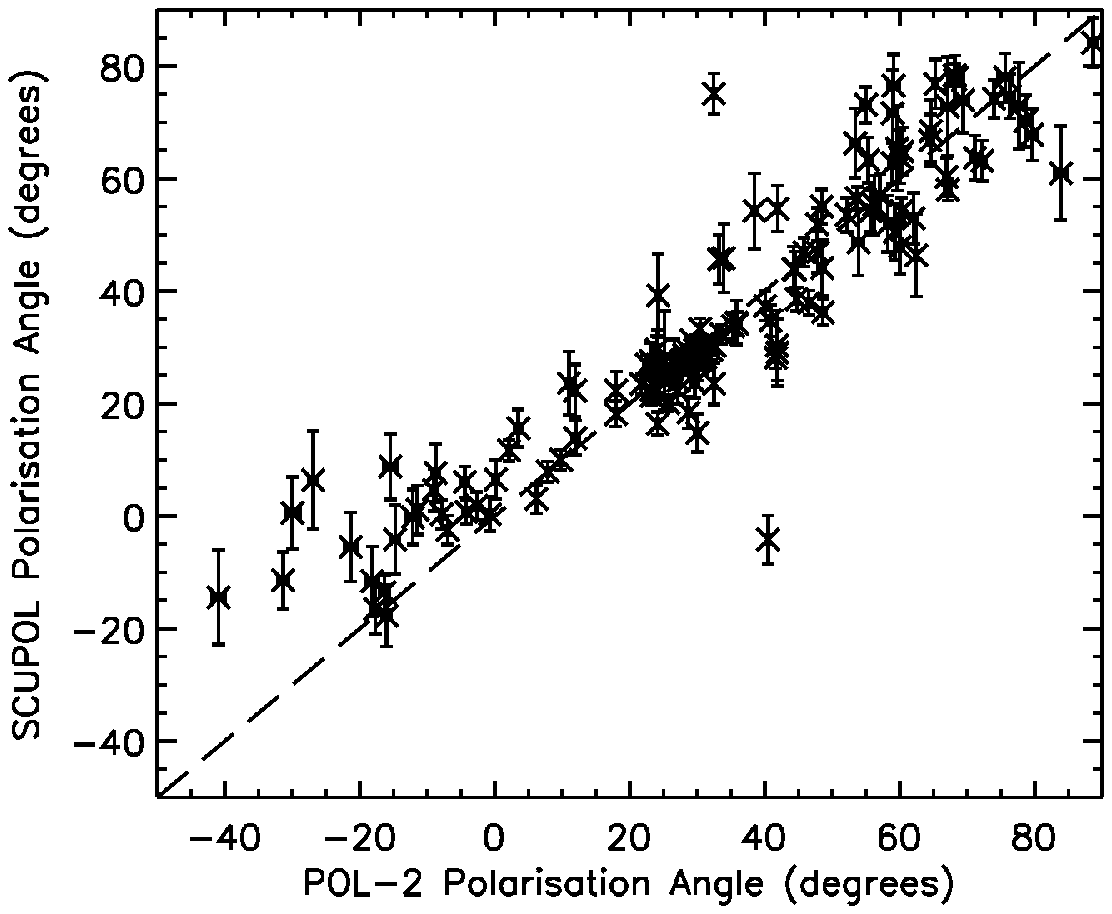}
\caption{Polarization angles at matched coordinates in the \mbox{POL-2} 
and SCUPOL maps. The dashed line shows the 1:1 line.}
\label{fig:pol2_scupol_ang}
\end{figure}

The similarity in the polarization angles of the \mbox{POL-2} and SCUPOL 
half-vectors is shown quantitatively in Figure~\ref{fig:pol2_scupol_ang}.  
The \mbox{POL-2} and SCUPOL polarization angles are plotted at positions 
matched to within one JCMT beam (14.1 arcsec). The two half-vector sets show 
correlated polarization angles, and in fact the \mbox{POL-2} and SCUPOL 
polarization angles are consistent with a 1:1 relationship.

\label{lastpage}

\end{document}